\documentclass[aps,prb,reprint,groupedaddress,footinbib]{revtex4-2}

\usepackage{amsmath}

\usepackage{graphicx}
\usepackage{verbatim}
\usepackage{amsfonts} 

\usepackage{mathbbol}
\usepackage{amssymb}   
\usepackage{color}
\newcommand{\be}{\begin{equation}}
\newcommand{\ee}{\end{equation}}
 
 \definecolor{BrickRed}{cmyk}{0,0.89,0.94,0.28}
\definecolor{MidnightBlue}{cmyk}{0.98,0.13,0,0.43}
\definecolor{DarkGreen}{rgb}{0,0.7,0.1}

\begin{document}

\title{Complex orientation dependence of Casimir-Polder interaction induced by curvature and optical properties of the surface and the surrounding medium}

\date{\today}

\author{Giuseppe Bimonte}
\affiliation{Dipartimento di Scienze Fisiche, Universit\`{a} di
Napoli Federico II, Complesso Universitario
di Monte S. Angelo,  Via Cintia, I-80126 Napoli, Italy}
\affiliation{INFN Sezione di Napoli, I-80126 Napoli, Italy}

\author{Thorsten Emig}
\affiliation{Laboratoire de Physique
Th\'eorique et Mod\`eles Statistiques, CNRS UMR 8626,
Universit\'e Paris-Saclay, 91405 Orsay cedex, France}

\begin{abstract}

We employ a multiple scattering expansion to systematically derive curvature corrections to the Casimir-Polder (CP) interaction between small an-isotropic particles and general magneto-dielectric surfaces. Our results, validated against exact solutions, reveal a complex, distance-dependent interplay between material properties and surface curvature in determining stable particle orientations. We demonstrate that even small surface curvature can induce or eliminate switches in the preferred orientation, a quantum effect that is diminished by thermal fluctuations. This work provides a crucial understanding of how to engineer nano-particle orientation through tune-able parameters, offering significant implications for micro- and nano-mechanical device design.

\end{abstract}


\maketitle

\section{Introduction}
\label{sec:intro}

Particle-surface interactions play a crucial role in numerous physical, chemical, and biological processes \cite{rev1,rev2,rev3,rev4,rev4bis,rev5,rev6}. While the underlying mechanisms can vary, many such interactions involve neutral particles influenced by fluctuating electromagnetic fields. Recent experimental investigations have focused on understanding how surface geometry, particularly microstructured surfaces, affects dispersion forces  ~\cite{exp1,exp2,exp3,exp4,exp5,bender}. Other notable examples include quantum friction \cite{P1997}, heat transfer between nanoparticles \cite{BG2010}, and critical Casimir forces in complex systems \cite{KHD2009,VED2013}.

This work delves into the realm of van der Waals and Casimir-Polder (CP) forces, arising from quantum and thermal fluctuations of the electromagnetic field. Traditionally studied for planar surfaces \cite{polder}, these forces exhibit significant modifications in the presence of curved geometries. The proximity force approximation (PFA) \cite{deri}, commonly used for short-distance interactions, often fails to capture curvature effects accurately. Furthermore, the non-additive nature of these forces can lead to intriguing phenomena for anisotropic particles  \cite{MAP2015}. For good conductors, the interaction between spheroids scales with the volumes of their enclosing spheres, rather than their actual volumes \cite{EGJK2009}.

A range of computational methods, including perturbative, scattering, and numerical techniques, can be applied to calculate complex Casimir-Polder forces, each with its own advantages and limitations.  Early 20th-century scattering techniques  \cite{sca1,sca2} are well-suited for distances significantly larger than the radii of curvature. However, their practical applicability is limited to highly symmetric bodies, such as spheres \cite{Marachevsky,Buhmann2} and cylinders \cite{galina,milton} , or a few perfectly conducting shapes \cite{PNAS}. Additionally, these methods face fundamental limitations in handling interlocked geometries due to the convergence issues of mode expansions.  Within the scattering framework, the Rayleigh expansion has been employed to study atom-nanograting interactions  \cite{Buhmann}. Perturbative methods ~\cite{messina} are applicable to surfaces with small, smooth corrugations.    For a small contrast between the permittivity of the body and the surrounding medium, the Casimir energy can be calculated using a Born series expansion. This approach expresses the energy as a power series in  the contrast, with each term given by an iterated volume integral over the bodies \cite{rev4bis, Buhmann:2006aa,fiedler}.

 An alternative approach that becomes exact in the limit of small particle-surface distances was introduced in \cite{kardar2014,Bimonte2015}. This method involves a systematic expansion of the potential in derivatives of the surface profile, enabling its application to general curved surfaces. A similar technique had been  previously employed  to study Casimir interactions between non-planar surfaces ~\cite{fosco2,bimonte3,bimonte4}, radiative heat transfer ~\cite{golyk}, and electrostatic forces between conductors ~\cite{fosco3}. For a recent review of the derivative expansion method, see \cite{fosco2024}.

Here, we reconsider the problem of the CP potential for an anisotropic particle near a curved surfaces, in the limit where the radii of curvature $R_j$, $j=1$, $2$ are much larger than the distance $d$ of the particle from the surface. Thanks to a recently developed surface formulation of a multiple scattering expansion (MSE) for Casimir and Casimir-Polder interactions  for magneto-dielectric materials \cite{bimonte2023,bimonte2023b,bimonte2023c,emig2024}, generalizing earlier work by Balian and Duplantier for perfect conductors \cite{BD1977},  the CP potential can be computed {\it without any approximation} to (at least) first order in $d/R_j$. At this order, we estimate an error of about $2\%$ of this expansion for $d/R_j=0.1$ by comparing to the known potential of a sphere.
This result represents a significant advance over earlier work \cite{kardar2014,Bimonte2015}. Firstly, the earlier derivative expansion relied on a reasonable {\it ansatz}, justified only a posteriori through a formal re-summation of a perturbative expansion for small in-plane momenta \cite{kardar2014,fosco2024}. The MSE provides a simple and direct curvature expansion, without requiring any prior assumptions. Secondly, and practically most important, the earlier approach in \cite{kardar2014,Bimonte2015} is limited to  perfect electric conductors (PEC), leaving material effects inaccessible. 
Here, we indeed find that combined material and curvature effects turn out be important for the orientation dependence of the CP potential for particles with anisotropic polarizability. 

The structure of the paper is as follows: Sec. \ref{sec:CP}  introduces the theoretical framework for the Casimir-Polder (CP) interaction between a polarizable particle and a magneto-dielectric body. It reviews the formula for the CP potential and discusses various approaches for its computation, including perturbative and scattering methods. It also provides a brief overview of the small-slope expansion proposed in previous works.
Sec. \ref{sec:MSE} details the multiple scattering expansion (MSE) and demonstrates how it can be used to prove the small-slope expansion derived in earlier works. It explains the concepts of surface operators and their connection to the free-space Green tensors. The section outlines the mathematical derivation of the small-slope expansion of the CP potential  by expanding the Green tensor in terms of surface curvature.
Sec. \ref{sec:valid}  assesses the validity range of the curvature expansion. It accomplishes this by comparing the approximate results from the expansion with the known exact solution for the CP potential of a spherical surface.
Sec. \ref{sec:orient} 
 investigates the orientation dependence of the CP force. It explores how factors like surface curvature, material properties, and thermal fluctuations affect the stable orientation of anisotropic particles. The analysis is performed for various materials, including Gold, Silicon, and Polystyrene, in different media.
Sec. \ref{sec:disc} concludes the paper with a discussion of the findings. It summarizes the work and its implications for controlling nano-particle orientation, highlighting the crucial role of material properties, surface curvature, and temperature.

\section{Casimir-Polder force between a polarizable particle and a magneto-dielectric non-planar surface}
\label{sec:CP}

The Casimir-Polder interaction describes the coupling of a small polarizable particle with the fluctuating electromagnetic field of a medium, which is modified by nearby magneto-dielectric bodies. We consider a particle embedded in a medium with electric and magnetic permittivities $\epsilon_0(\omega)$ and $\mu_0(\omega)$, placed at a minimum distance $d$ from a magneto-dielectric body with permittivities $\epsilon_1(\omega)$ and $\mu_1(\omega)$.
If the particle has a 
frequency-dependent electric polarizability tensor ${\alpha}(\omega)$  (for simplicity, we assume that the magnetic polarizability  is negligible), the classical energy of its induced dipole is given by the following formula:
\begin{equation}
\label{eq:inducedDipole}
U_\text{cl} = -\frac{1}{2} \sum_{i,j=1}^{3} \alpha_{ij} E_i E_j  \, .
\end{equation}
Using the fluctuation-dissipation theorem, this expression is averaged over EM field fluctuations.  After a divergent contribution from empty space is removed, the CP potential is expressed by the formula:
\begin{eqnarray}
\label{eq:CPenergy}
\!\!\!\!\!\!U &=& - 4\pi k_B T \sideset{}{'}\sum_{n=0}^\infty  \!\!\kappa_n \!\!\!\sum_{i,j=1}^{3}  \alpha_{ij} ({\rm i} \, \xi_n) \mathbb{\Gamma}^{(EE)}_{ij}({\bf r}_0,{\bf r}_0;\kappa_n) .
\end{eqnarray}
Here  $\xi_n=2\pi k_B T n/\hbar$ are the Matsubara frequencies,   $\kappa_n=\xi_n/c$,  the prime in the sum indicates that the  $n=0$ term is taken with weight one-half and ${\bf r}_0$  denotes  the particle's  position. The scattering Green tensor $\mathbb{\Gamma}({\bf r},{\bf r}') =\mathbb{G}({\bf r},{\bf r}') -\mathbb{G}_0({\bf r},{\bf r}') $ is defined as the difference between the body Green tensor  
$\mathbb{G}({\bf r},{\bf r}') $ and the empty space Green tensor $\mathbb{G}_0({\bf r},{\bf r}') $  for the homogeneous medium. The formula yields the exact CP potential for a limited number of highly symmetric bodies where the scattering Green tensor $\mathbb{\Gamma}$ is known. These include a plane \cite{rev4}, a sphere \cite{Marachevsky,Buhmann2}, a cylinder \cite{galina,milton}, or a few perfectly conducting shapes \cite{PNAS}. Beyond these special geometries, a variety of approximate approaches have been developed to compute $\mathbb{\Gamma}$ for more general shapes. For example, a perturbative approach can be used for planar surfaces with smooth, low-amplitude corrugations \cite{messina}. Alternatively, the Green tensor   $\mathbb{\Gamma}({\bf r},{\bf r}')$  can be expressed as a multiple scattering expansion using a Born series \cite{rev4bis,Buhmann:2006aa,fiedler}.  This expansion represents the tensor as a power series of the contrast parameters, $\delta \epsilon({\bf r},\omega)=\epsilon_1({\bf r},\omega)-\epsilon_0({\bf r},\omega)$ and $\delta \mu({\bf r},\omega)=\mu_1({\bf r},\omega)-\mu_0({\bf r},\omega)$, with terms given by iterated volume integrals over the bodies.

\begin{figure}[h]
\includegraphics [width=.9\columnwidth]{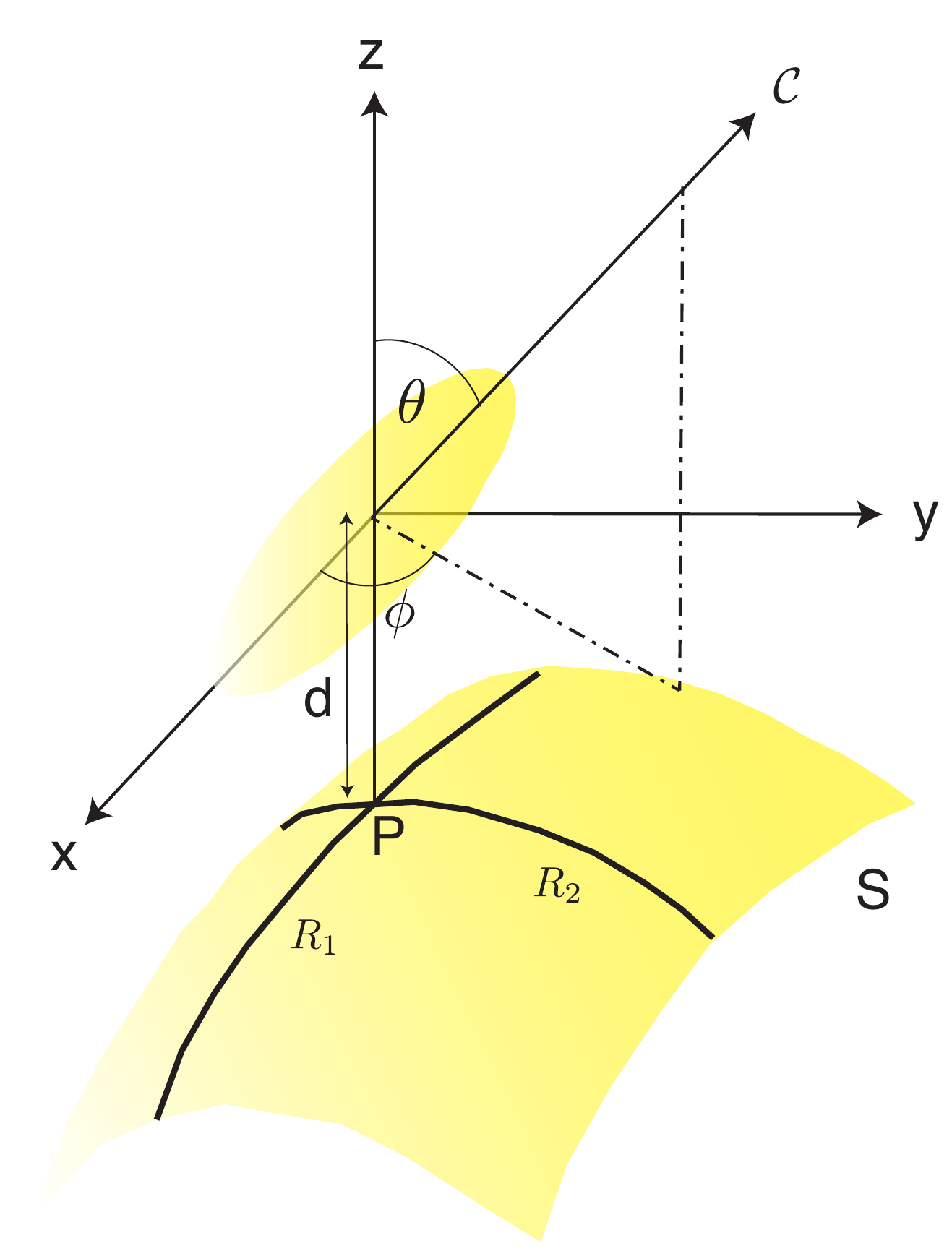} 
\caption{\label{orient}   This figure illustrates the configuration of a small ellipsoidal particle positioned near a gently curved surface. The solid curves on surface $S$ at point $P$ represent the principal directions. The corresponding local radii of curvature, $R_1$ and $R_2$ are positive when the surface curves towards the particle and negative when it curves away. }
\end{figure}

An approximation method that is valid for surfaces with ${\it small\;slope}$ was proposed in \cite{kardar2014,Bimonte2015}. Because this approach is closely related to our work, we will briefly review it here. 
We define a local coordinate system to describe the geometry. We place the particle at the origin of a plane, $\Sigma_1$, that is orthogonal to the vector connecting it to the closest point, $P$, on the body's surface, $S$. We align this connecting vector with the $z$-axis, so the particle is at $z=0$ and the closest point on the surface is at $z=-d$. The surface $S$ is assumed to have a smooth profile given by the function $z=-d+h({\bf u})$, where  the two dimensional vector  ${\bf u}=(x,y) \equiv (u_1,u_2)$ spans the plane $\Sigma_1$ (see Fig. \ref{orient}). 
Given that the CP interaction falls off rapidly with distance, the core idea of \cite{Bimonte2015} is to assume that the potential $U$ is dominated by a small neighborhood of the closest point, $P$, on the surface $S$. This physically plausible assumption suggests that for small separations $d$, the potential can be expanded as a series in an increasing number of derivatives of the height profile $h$, all evaluated at position $P$.

When the components of the polarizability tensor are expressed relative to the principal directions of surface curvature at the point $P$, an expansion of all surface operators to linear order in the second derivatives of the surface profile $h({\bf u})$ yields the CP potential at temperature T (see Sec.~III below),
\begin{widetext}
\be
\label{derexpa}
  U = -  \frac{k_B T}{d^3} \sideset{}{'}\sum_{n=0}^\infty \left\{\beta^{(0)}_{1} \alpha_\perp + \beta^{(0)}_{2} \alpha_{zz} + d \, \left[ (\beta^{(2)}_{1} \alpha_\perp + \beta^{(2)}_{2} \alpha_{zz}) \nabla^2 h+
 \beta^{(2)}_{3}  \left(\partial_i \partial_j h - \frac{1}{2} \nabla^2 h \delta_{ij}\right) \alpha_{ij}\right] \right\} \;, 
 \ee
 \end{widetext}
 where  $\alpha_\perp=\alpha_{xx}+\alpha_{yy}$,  and it is understood that all derivatives of $h$ are evaluated at the position $P$, i.e.,
for ${\bf u}=(0,0)$. The dimensionless coefficients  $\beta^{(p)}_q$ are   functions  of $\bar{\kappa}_n=\kappa_n d$ and the electric and magnetic permittivities of both the surface  ($\epsilon_1(i\xi_n)$, $\mu_1(i\xi_n)$) and the embedding medium ($\epsilon_0(i\xi_n)$, $\mu_0(i\xi_n)$). In the limit $T\to 0$, the summation is replaced by an integration according to $k_B T \sum\nolimits_{n=0}^{'\infty} \to \frac{\hbar c}{d} \int_0^\infty \frac{d\tilde\kappa}{2\pi}$. Note that the same expression for $U$ applies to particles with magnetic polarizability but with different functions $\beta^{(p)}_q$. 
 The derivative expansion Eq.~(\ref{derexpa})  was first proposed as an ${\it ansatz}$ in \cite{kardar2014,Bimonte2015}. However, it can also be formally derived by re-summing the perturbative series for the potential at small in-plane momenta ${\bf k}$ \cite{kardar2014}.
 A clearer understanding of Eq.~(\ref{derexpa})  emerges from the local expansion  $h=u_1^2/(2 R_1)+ u_2^2/(2 R_2)+\ldots$,  where $R_1$ and $R_2$ are the radii of curvature at $P$. 
 In this coordinate system, the derivative expansion of $U$ can be written as:
\begin{widetext}
\be
U  =-\frac{\hbar c}{d^4} \int_0^{\infty} \frac{d \xi}{2 \pi}\left\{\beta^{(0)}_1 \alpha_\perp + \beta^{(0)}_2 \alpha_{zz} + \left(\frac{d}{R_1}+\frac{d}{R_2} \right) (\beta^{(2)}_1 \alpha_\perp + \beta^{(2)}_2 \alpha_{zz})+
 \frac{\beta^{(2)}_3}{2}  \left(\frac{d}{R_1}-\frac{d}{R_2}\right) (\alpha_{xx}-\alpha_{yy}) \right\} 
\;.\label{derexpa2}
\ee
\end{widetext}
This expression shows that the terms linear in the second derivatives of $h$ in Eq.~(\ref{derexpa})  actually represent curvature corrections.
 The coefficients $\beta^{(p)}_q$
for a perfectly conducting surface in vacuum were computed in \cite{kardar2014}.  The authors achieved this by matching the derivative expansion with the perturbative expansion of $U$ in their shared domain of validity.

\section{The multiple scattering expansion and its small slope expansion}
\label{sec:MSE}

In this section, we show that the small slope expansion Equation (\ref{derexpa}) can be proven by a straightforward computation. The proof begins with the exact representation of the scattering Green tensor  $\mathbb{\Gamma}({\bf r}_0,{\bf r}_0)$   provided by the  MSE introduced in \cite{bimonte2023,bimonte2023b,bimonte2023c,emig2024},
\begin{widetext}
\begin{equation}
\label{eq:GammaEE}
\mathbb{\Gamma}^{(EE)}({\bf r}_0,{\bf r}_0;{\kappa}_n) =\int_S d\sigma_{\bf u} \int_S d\sigma_{{\bf u}'}\, \!\!\!\!\sum_{p,q\in \{E,H\}}\!\!\!\!\! \mathbb{G}_0^{(Ep)}({\bf r}_0,{\bf s}({\bf u})) \left[(\mathbb{I}-\mathbb{K})^{-1}\right]^{(pq)}({\bf s}({\bf u}),{\bf s}({\bf u}')) \mathbb{M}^{(qE)}({\bf s}({\bf u}'),{\bf r}_0)\,,
\end{equation}
\end{widetext}
The surface area element on $S$ is given by $d\sigma_{\bf u}=\sqrt{1+(\partial h/\partial_{u_1})^2+(\partial h/\partial_{u_2})^2}$. 
The surface operators $\mathbb{K}$ are defined for positions ${\bf s}$ and ${\bf s}'$  on the surface. Here we parameterized these positions as   ${\bf s}({\bf u})={\bf u}- d  \hat{\bf z}+h({\bf u})\hat{\bf z}$.  The operators $\mathbb{K}$  are expressed in terms of the  $\it{free}$ Green tensor $\mathbb{G}_0$  for $\it{homogeneous}$ space (for  a medium  with permittivities $\epsilon_0$, $\mu_0$)  and $\mathbb{G}_1$ (for  a medium  with permittivities $\epsilon_1$, $\mu_1$)   (see Appendix  \ref{sec:green} for the explicit expressions of $\mathbb{G}_0$ and $\mathbb{G}_1$),
\begin{widetext}
\begin{equation}
\begin{aligned}
\label{eq:App_K}
\mathbb{K}^{(EE)}({\bf s},{\bf s}') & = \frac{2}{\mu_0+\mu_1} {\bf n}({\bf s}) \times \left[ \mu_0 \mathbb{G}_0^{(HE)}({\bf s},{\bf s}') - \mu_1 \mathbb{G}_1^{(HE)}({\bf s},{\bf s}')\right] \;,\\
\mathbb{K}^{(HH)}({\bf s},{\bf s}') & = \frac{2}{\epsilon_0+\epsilon_1} {\bf n}({\bf s}) \times \left[ -\epsilon_0 \mathbb{G}_0^{(EH)}({\bf s},{\bf s}') +  \epsilon_1 \mathbb{G}_1^{(EH)}({\bf s},{\bf s}')\right] \;,\\
\mathbb{K}^{(EH)}({\bf s},{\bf s}') & = \frac{2}{\mu_0+\mu_1} {\bf n}({\bf s}) \times \left[ \mu_0 \mathbb{G}_0^{(HH)}({\bf s},{\bf s}') - \mu_1 \mathbb{G}_1^{(HH)}({\bf s},{\bf s}')\right] \;,\\
\mathbb{K}^{(HE)}({\bf s},{\bf s}') & = \frac{2}{\epsilon_0+\epsilon_1} {\bf n}({\bf s}) \times \left[ - \epsilon_0 \mathbb{G}_0^{(EE)}({\bf s},{\bf s}') +  \epsilon_1 \mathbb{G}_1^{(EE)}({\bf s},{\bf s}')\right] \;.
\end{aligned} 
\end{equation}
Here,  ${\bf n}({\bf s})$ is the outward-pointing surface normal vector. In our parametrization, it can be written as
${\bf n}=\tilde {\bf n}/\sqrt{1+(\partial h/\partial_{u_1})^2+(\partial h/\partial_{u_2})^2}$,  with $\tilde {\bf n}=\hat{\bf z} - \hat{\bf x} \partial h /\partial u_1 - \hat{\bf y} \partial h /\partial u_2$.  The  operators $\mathbb{M}$  are instead defined for position ${\bf s}$ on the surface, and ${\bf r}$ in the medium outside $S$,
\begin{equation}
\begin{aligned}
\label{eq:App_M}
\mathbb{M}^{(EE)}({\bf s},{\bf r}) & = \frac{2\mu_0}{\mu_0+\mu_1} {\bf n}({\bf s}) \times \mathbb{G}_0^{(HE)}({\bf s},{\bf r})\,, \\
\mathbb{M}^{(EH)}({\bf s},{\bf r})  & = \frac{2\mu_0}{\mu_0+\mu_1} {\bf n}({\bf s}) \times \mathbb{G}_0^{(HH)}({\bf s},{\bf r})\;,\\
\mathbb{M}^{(HE)}({\bf s},{\bf r}) & = - \frac{2\epsilon_0}{\epsilon_0+\epsilon_1} {\bf n}({\bf s}) \times \mathbb{G}_0^{(EE)}({\bf s},{\bf r})\,, \\
\mathbb{M}^{(HH)}({\bf s},{\bf r})  &= - \frac{2\epsilon_0}{\epsilon_0+\epsilon_1} {\bf n}({\bf s}) \times \mathbb{G}_0^{(EH)}({\bf s},{\bf r}) \, .
\end{aligned}
\end{equation}
\end{widetext}
 The representation of the scattering Green tensor $\mathbb{\Gamma}({\bf r},{\bf r}')$ in terms of surface operators, as shown in Eq. (\ref{eq:GammaEE}), provides a direct way to derive the small-slope expansion of the potential $U$.
 The key to this derivation is a crucial observation: at imaginary frequencies $\xi$, the free-space Green tensors,  $\mathbb{G}_0$  and $\mathbb{G}_1$, decay exponentially with the separation between their two arguments. This property immediately implies that the points ${\bf s}$ on the surface $S$ that contribute most significantly to the MSE of the potential $U$ are those clustered within a small neighborhood of the closest point, $P$, to the particle.
This finding justifies the use of a Taylor expansion for the surface parameterization ${\bf s}({\bf u})$ around the point where ${\bf u} = {\bf 0}$, which corresponds to the location of $P$. 
When the height profile $h$ is expanded to second order, we find:
\begin{eqnarray}
\label{eq:App_surface}
&&{\bf s}(u_1,u_2) =u_1\hat{\bf x}+u_2\hat{\bf y}-d \hat{\bf z}
+h(u_1,u_2)\hat{\bf z} \nonumber \\
&&=u_1\hat{\bf x}+u_2\hat{\bf y}-d \hat{\bf z}
 +\frac{1}{2}h_{\alpha\beta}u_\alpha u_\beta \hat{\bf z} +\ldots
\end{eqnarray}
Here, we've used the fact that both $h$ and its first derivatives vanish at the origin  $u_1=u_2=0$,   and we defined $h_{\alpha\beta}$ as the second derivatives of $h$ at $u_1=u_2=0$: $h_{\alpha\beta}=\partial^2 h/\partial_{u_\alpha}\partial_{u_\beta}(0,0)$. 
 The outward surface normal vector   ${\bf n}=\tilde {\bf n}/\sqrt{1+(\partial h/\partial_{u_1})^2+(\partial h/\partial_{u_2})^2}$ to lowest order in curvature is given by $\tilde {\bf n}=\hat{\bf z} - \hat{\bf x} h_{1\alpha} u_\alpha - \hat{\bf y} h_{2\alpha} u_\alpha$.
We note that the normalizing square root can be ignored when evaluating Eq.~(\ref{eq:GammaEE}) as it cancels against the surface element from the surface integrals. 

 Next we expand all operators to linear order in $h_{\alpha\beta}$. We denote by $\overset{\text{\tiny 0}}{\mathbb{G}}_\sigma$, $\overset{\text{\tiny 0}}{\mathbb{K}}$ and $\overset{\text{\tiny 0}}{\mathbb{M}}$ the operators for $h_{\alpha\beta}=0$ and by $\overset{\text{\tiny 1}}{\mathbb{G}}_\sigma$, $\overset{\text{\tiny 1}}{\mathbb{K}}$ and $\overset{\text{\tiny 1}}{\mathbb{M}}$ their linear order $\sim h_{\alpha\beta}$. The zeroth order terms $\overset{\text{\tiny 0}}{\mathbb{G}}_\sigma$, $\overset{\text{\tiny 0}}{\mathbb{K}}$ and $\overset{\text{\tiny 0}}{\mathbb{M}}$ of course coincide with the  corresponding operators for the planar surface of equation $z=-d$.
The zeroth and first order terms of the expansion of the electric scattering Green tensor in surface curvature can be expressed in compact notation as
\begin{widetext}
\begin{equation}
\begin{aligned}
\label{eq:App_Gamma_exp}
\overset{\text{\tiny 0}}{\mathbb{\Gamma}}^{(EE)} & = \begin{pmatrix} \overset{\text{\tiny 0}}{\mathbb{G}}_0^{(EE)} \\ \overset{\text{\tiny 0}}{\mathbb{G}}_0^{(EH)} \end{pmatrix} {\mathbb O} \begin{pmatrix} \overset{\text{\tiny 0}}{\mathbb{M}}^{(EE)} \\ \overset{\text{\tiny 0}}{\mathbb{M}}^{(EH)} \end{pmatrix}  \\ 
\overset{\text{\tiny 1}}{\mathbb{\Gamma}}^{(EE)} & = \begin{pmatrix} \overset{\text{\tiny 1}}{\mathbb{G}}_0^{(EE)} \\ \overset{\text{\tiny 1}}{\mathbb{G}}_0^{(EH)} \end{pmatrix} {\mathbb O} \begin{pmatrix} \overset{\text{\tiny 0}}{\mathbb{M}}^{(EE)} \\ \overset{\text{\tiny 0}}{\mathbb{M}}^{(EH)} \end{pmatrix} 
+ \begin{pmatrix} \overset{\text{\tiny 0}}{\mathbb{G}}_0^{(EE)} \\ \overset{\text{\tiny 0}}{\mathbb{G}}_0^{(EH)} \end{pmatrix} {\mathbb O} \begin{pmatrix} \overset{\text{\tiny 1}}{\mathbb{M}}^{(EE)} \\ \overset{\text{\tiny 1}}{\mathbb{M}}^{(EH)} \end{pmatrix}
+ \begin{pmatrix} \overset{\text{\tiny 0}}{\mathbb{G}}_0^{(EE)} \\ \overset{\text{\tiny 0}}{\mathbb{G}}_0^{(EH)} \end{pmatrix} {\mathbb O} \overset{\text{\tiny 1}}{\mathbb{K}} {\mathbb O} \begin{pmatrix} \overset{\text{\tiny 0}}{\mathbb{M}}^{(EE)} \\ \overset{\text{\tiny 0}}{\mathbb{M}}^{(EH)}\;. \end{pmatrix}
\end{aligned}
\end{equation}
Here
\begin{equation}
\begin{aligned}
\label{eq:App_O2}
\mathbb{O} &= \left( \mathbb{1} - \overset{\text{\tiny 0 }}{\mathbb{K}}\right)^{-1} = 
\begin{pmatrix}
\mathbb{R} &\,\,\, \mathbb{R} \overset{\text{\tiny 0 }}{\mathbb{K}}^{(EH)} \\
\overset{\text{\tiny 0 }}{\mathbb{K}}^{(HE)}\mathbb{R} &\,\,\, \mathbb{R}
\end{pmatrix} \\
\mathbb{R} &= \left( \mathbb{1} - \overset{\text{\tiny 0 }}{\mathbb{K}}^{(EH)}\overset{\text{\tiny 0 }}{\mathbb{K}}^{(HE)} \right)^{-1}
\end{aligned}
\end{equation}
\end{widetext}
where we used that $\overset{\text{\tiny 0 }}{\mathbb{K}}^{(EE)}=0$, $\overset{\text{\tiny 0 }}{\mathbb{K}}^{(HH)}=0$.
To evaluate the above expressions, we used the Fourier representation of $\mathbb{G}_0$  and $\mathbb{G}_1$ and carried out analytically the
lengthy integrations over the wave vector along the $\hat{\bf z}$ direction and the angle of the in-plane $\hat{\bf x}$-$\hat{\bf y}$ wave vector. This yields to zeroth order  the diagonal tensor $4\pi \kappa \, \overset{\text{\tiny 0}}{\mathbb{\Gamma}}^{(EE)}=d^{-3}\,\text{diag}(\beta_1^{(0)},\beta_1^{(0)},\beta_2^{(0)})$ with
\begin{widetext}
\begin{equation}
\begin{aligned}
\label{eq:betas_O}
\beta_1^{(0)} &= \int_0^\infty \!\!  \bar{k} d\bar{k} \, \frac{e^{-2 s_0} \left(2 \bar{\kappa} ^4 \mu_0^2 \epsilon_0^2 s_1 (\mu_0 \epsilon_1-\mu_1 \epsilon_0)+3 \bar{\kappa} ^2 \bar{k}^2 \mu_0
   \epsilon_0 s_1 (\mu_0 \epsilon_1-\mu_1 \epsilon_0)+\bar{k}^4 \left(\epsilon_1 \left(\mu_1 s_0+\mu_0 s_1 \right)-\epsilon_0 \left(\mu_0 s_0+\mu_1 s_1\right)\right)\right)}{2 \epsilon_0 s_0^2 \left(\mu_1 \epsilon_0 \left(s_0 s_1 +2 \tilde \kappa ^2 \mu_0 \epsilon_1\right)+\mu_0 \epsilon_1 s_0 s_1+\bar{k}^2 (\mu_0 \epsilon_0+\mu_1
   \epsilon_1)\right)} \\
\beta_2^{(0)} &= \int_0^\infty \!\!  \bar{k} d\bar{k} \, \frac{e^{-2 s_0} \left(\bar{k}^4 \left(\mu _1 \epsilon _1-\mu _0 \epsilon _0\right)+\bar{k}^2 s_0 s_1 \left(\mu _0 \epsilon _1-\mu _1 \epsilon
   _0\right)\right)}{\epsilon _0 s_0 \left(\bar{k}^2 \left(\mu _0 \epsilon _0+\mu _1 \epsilon _1\right)+\mu _1 \epsilon _0
   \left(s_0 s_1+2 \tilde \kappa ^2 \mu _0 \epsilon _1\right)+\mu _0 s_0 s_1 \epsilon _1\right)}
\end{aligned}
\end{equation}
\end{widetext}
where $s_\sigma=\sqrt{\bar{k}^2+ \mu_\sigma \epsilon_\sigma \bar{\kappa}^2}$. Here we have introduced the rescaled quantities $\bar{k} = d k$, $\bar{\kappa} = d\kappa$ so that $\beta_1^{(0)}$ and $\beta_2^{(0)}$ are dimensionless. 
The next order, which is linear in curvature, assumes the form
\begin{widetext}
\begin{equation}
\label{eq:Gamma1_O}
4\pi \kappa \, \overset{\text{\tiny 1}}{\mathbb{\Gamma}}^{(EE)}= \frac{1}{d^2} \begin{pmatrix}
\left( \beta_1^{(2)} + \frac{1}{2} \beta_3^{(2)} \right) h_{11} + \left( \beta_1^{(2)} -\frac{1}{2} \beta_3^{(2)} \right) h_{22} & \beta_3^{(2)} h_{12} & 0 \\
\beta_3^{(2)} h_{12} & \left( \beta_1^{(2)} - \frac{1}{2} \beta_3^{(2)} \right) h_{11} + \left( \beta_1^{(2)} +\frac{1}{2} \beta_3^{(2)} \right) h_{22} & 0 \\
0 & 0 & \beta_2^{(2)} \left( h_{11} + h_{22} \right)\, .\\
\end{pmatrix}
\end{equation}
\end{widetext}
Here the $\beta$ coefficients are given by $k$-integrals similar to those in Eq.~(\ref{eq:betas_O}) but their  expressions are too long to be shown here. Instead they are provided in a supplementary Mathematica notebook \cite{Mathematica}.
When the expressions for $\overset{\text{\tiny 0}}{\mathbb{\Gamma}}^{(EE)}$ and $\overset{\text{\tiny 1}}{\mathbb{\Gamma}}^{(EE)}$ are substituted into 
Eq.~(\ref{eq:CPenergy}) one obtains the small slope expansion  CP potential in 
Eq.~(\ref{derexpa}).
 
\section {Validity range of curvature expansion}
\label{sec:valid}

\begin{figure}[h]
\includegraphics [width=.9\columnwidth]{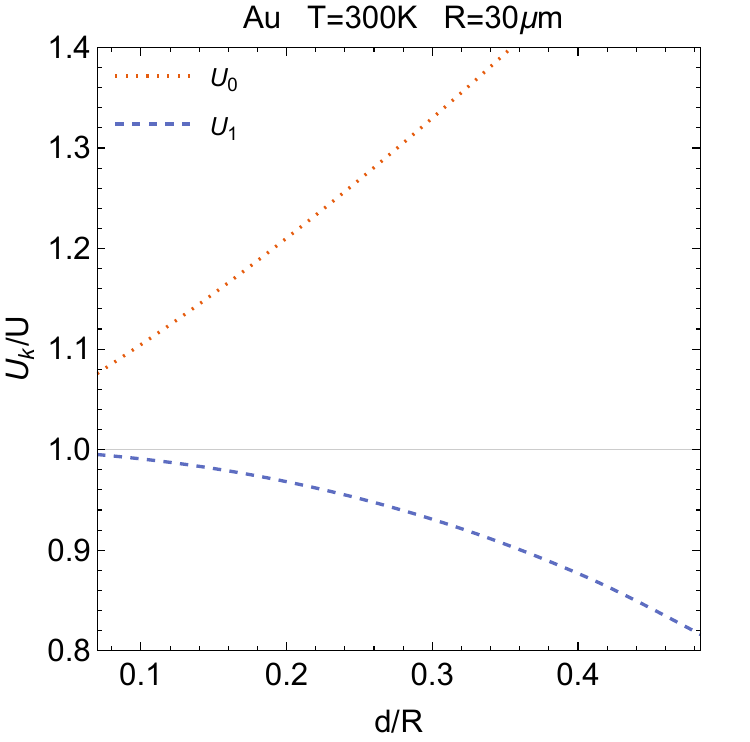} 
\caption{\label{approxAu}   Ratio of the curvature expansion of the CP potential $U_k$ up to the k-th power of $d/R$ to the exact potential $U_\text{sphere}$ for an Au sphere of radius $R=30\mu$m at  $T=300K$ versus the normalized distance $d/R$.}
\end{figure}

The validity range of the small curvature expansion of Eq.~(\ref{derexpa}) can be estimated as the exact CP potential  for the special case of a spherical surface is known from a scattering approach \cite{Buhmann2}. For simplicity, here we consider a particle with isotropic, frequency independent polarizability $\alpha_{ij} =\alpha \, \delta_{ij}$ at distance $d$ from a dielectric sphere of radius $R$ in vacuum ($\epsilon_0=\mu_0=1$). 
In the partial wave scattering approach, the CP potential can be written as \cite{Marachevsky,Buhmann2}
\begin{widetext}
\begin{eqnarray}
\label{eq:E_sphere}
{U}_\text{sphere} =\frac{k_B T}{a^2} \sideset{}{'}\sum_{n=0}^\infty  \kappa_n \,  \alpha({\rm i} \, \xi_n) \sum_{l=1}^{\infty} (2 l +1)   \left\{T^{\rm HH}_{l}  ({\rm i} \, \xi_n)  {\cal K}_l^2 (\kappa_n a) -
T^{\rm EE}_l ({\rm i} \, \xi_n) \left[ {\cal K}_l^{'2} (\kappa_n a) + \frac{l(l+1)}{\kappa_n^2 a^2} {\cal K}_l^2 (\kappa_n a)  \right]\right\}\nonumber \;,
\end{eqnarray}
where $a=R+d$,  $l$ is the multipole index,  ${\cal K}'_l(x)= d {\cal K}_l /dx$,  ${\cal K}_l (x) = x k_l(x)$,  $k_l(x) = \sqrt{\frac{2}{\pi x}} K_{l+1/2}(x)$  is the modified spherical Bessel function of the third kind, and $T^{{\rm HH}}_l , T^{\rm EE}_l $ are the T-matrix elements (Mie coefficients) of the sphere,
\begin{eqnarray}
T^{\rm HH}_{l} ({\rm i} \xi)&=& \frac{ \sqrt{\mu/\epsilon}\, {\cal I}_l (\sqrt{\epsilon \mu}\kappa R) \; {\cal I}'_l (\kappa R) -   {\cal I}'_l (\sqrt{\epsilon \mu}\kappa R)  \;{\cal I}_l (\kappa R)  }
{{\cal K}_l (\kappa R) \; {\cal I}'_l ( \sqrt{\epsilon \mu}\kappa R) - \sqrt{\mu/\epsilon} \; {\cal I}_l (\sqrt{\epsilon \mu}\kappa R) \; {\cal K}'_l (\kappa R) } \;,\\
T^{\rm EE}_{l} ({\rm i} \xi)&=& \frac{ \sqrt{\epsilon/\mu}\, {\cal I}_l (\sqrt{\epsilon \mu}\kappa R) \; {\cal I}'_l (\kappa R) -   {\cal I}'_l (\sqrt{\epsilon \mu}\kappa R)  \;{\cal I}_l (\kappa R)  }
{{\cal K}_l (\kappa R) \; {\cal I}'_l ( \sqrt{ \mu/\epsilon}\kappa R) - \sqrt{\epsilon/\mu} \; {\cal I}_l (\sqrt{\epsilon \mu}\kappa R) \; {\cal K}'_l (\kappa R) } \;,\\
\end{eqnarray} 
\end{widetext}
where $\xi=\kappa c$, ${\cal I}_l (x) = x i_l(x)$ and ${\cal I}'_l(x)= d {\cal I}_l /dx$  with  $i_l(x) = \sqrt{\frac{\pi}{2 x}} I_{l+1/2}(x)$   the modified spherical Bessel function of the first kind. Fig.~\ref{approxAu} displays the ratios of the approximate CP potential $U_k$ (truncated at the $k$-th power of $d/R$) to the exact potential $U_\text{sphere}$ for a Au sphere with a radius of $R=30\mu$m at $T=300K$, as a function of the ratio $d/R$. The curvature expansion is remarkably accurate: For the smallest considered $d/R$ value of $0.07$, the flat surface approximation $U_0$ yields a relative error of $7.6 \%$. When the leading curvature correction is included, for $U_1$ the error reduces to $0.5\%$.

 \section{Orientation dependence}
\label{sec:orient} 

\begin{figure}[h]
\includegraphics [width=.9\columnwidth]{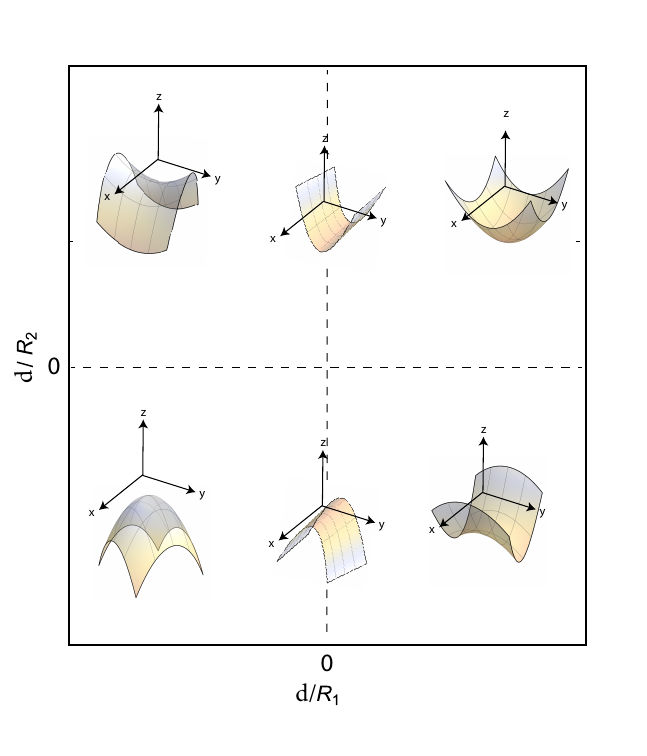} 
\caption{\label{shapes}  Typical surface shapes  corresponding to different combinations of $d/R_1$ and $d/R_2$. The coordinate frames indicate the position and orientation of the particle.}
\end{figure}

In this section we investigate the shape and orientation dependence of the Casimir-Polder force.
We assume that the particle has rotational symmetry about one axis ($3$-axis) so that the polarizability tensor is diagonal with $\tilde\alpha = \text{diag}(\tilde\alpha_\perp/2,\tilde\alpha_\perp/2,\tilde\alpha_3)$ relative to the principal axes attached to the particle. The corresponding  cartesian components of the polarization tensor in the $(x,y,z)$ coordinate system then read
\begin{eqnarray}
&&\alpha_{\perp}= \left[3 \tilde\alpha_\perp+ 2  \tilde\alpha_{3} - 2\sigma \cos(2 \theta) \right]\;,\\
&&\alpha_{zz}= \left[\tilde\alpha_\perp+ 2  \tilde\alpha_{3} + 2\sigma \cos(2 \theta) \right]\;,\\
&&\alpha_{xx}-\alpha_{yy}=  \sigma \cos(2 \phi)\,\sin^2 \theta\;,
\end{eqnarray}
where  $\sigma = \tilde\alpha_3 - \tilde\alpha_\perp/2$ measures the anisotropy of the particle's (frequency dependent) polarizability.
Upon substituting the above expressions into Eq. (\ref{derexpa2}), the orientation-dependent component of the potential is given by:
\begin{widetext}
\begin{equation}
\label{eq:U_orientation}
U_{o}(d,\theta,\phi) = -\frac{k_B T}{2 d^3} \, \sideset{}{'}\sum_{n=0}^\infty \!\!\sigma(i \xi_n) \left[\left\{ \Delta\beta^{(0)}(\bar{\kappa}_n) + \Delta\beta^{(2)}(\bar{\kappa}_n) \, \left(\frac{d}{R_1}+ \frac{d}{R_2}\right) \right\} \cos(2\theta) + \,\beta^{(2)}_3(\bar{\kappa}_n) \, \left(\frac{d}{R_1}- \frac{d}{R_2}\right) \cos(2\phi) \sin^2 \theta \right] \;.
\end{equation}
\end{widetext}
Here, we define $\Delta\beta^{(0)}=\beta^{(0)}_2-\beta^{(0)}_1$ and $\Delta\beta^{(2)}=\beta^{(2)}_2-\beta^{(2)}_1$, where $\beta^{(p)}_q$ are the coefficients introduced in Eq.~(\ref{derexpa}). The angle $\theta$ is formed by the particle's $3$-axis and the $z$-axis, and $\phi$ is the polar angle of the $3$-axis projected onto the $xy$-plane (see Fig. \ref{orient}).  The dimensionless functions $\Delta\beta^{(0)}$, $\Delta\beta^{(2)}$, $\beta^{(2)}_3$ depend on $\bar{\kappa}_n$ and on  the electric and magnetic permittivities of the surface ($\epsilon_1(i\xi_n)$, $\mu_1(i\xi_n)$) and the embedding medium ($\epsilon_0(i\xi_n)$, $\mu_0(i\xi_n)$). 
Note that Eq.~(\ref{eq:U_orientation}) requires no assumption about the distance $d$ relative to all material and medium related length scales. 
If $R_1\neq R_2$ and the sum of $\beta^{(2)}_3(\bar{\kappa}_n)\neq 0$ it is easy to verify that the potential $U_{o}$ has a {\it unique} minimum which corresponds to an orientation of the particle along one of the three principal directions of the surface ($x,y,z$). For $R_1=R_2$ or the sum of $\beta^{(2)}_3(\bar{\kappa}_n)=0$ the particle is oriented either along the $z$-axis or along any direction in the $xy$ (tangential) plane. To provide a clearer understanding of the particle's orientation relative to the curved surface, we show in Fig. \ref{shapes}  the typical surface shapes for positive and negative radii of curvature, along with the coordinate frames for the position and orientation of the particle.  
It is known that for a PEC {\it flat} surface there is no orientation dependence of the CP potential at zero temperature for dipolar particles with frequency independent polarizabilities.  This can be seen from Eq.~(\ref{eq:U_orientation}) since for $T=0$ the sum is replaced by an integral and $\int_0^\infty d\kappa \,\Delta\beta^{(0)}(\kappa)=0$. 
For general surface materials, frequency dependent polarizabilities, or finite temperature, however, the term proportional to $\Delta\beta^{(0)}$ is non-zero in general. Hence in the stable orientation the symmetry axis of the particle is either oriented perpendicular to the surface or can rotate freely in the surface plane. 
For all studied materials, we find that the preferred orientation does not depend on the distance $d$ of the particle from the surface if the surface is placed in vacuum. But interestingly, as we show here, the preferred orientation can depend on $d$, if the medium in which  the surface is placed has suitable optical properties. 

For a {\it curved} surface the orientation dependence becomes more complex.  
Remarkably, we find that small surface curvature can induce or eliminate switches in stable particle orientation, with the switching distance depending on material properties and magnitude of curvature. These switches are a quantum effect which is gradually washed out by thermal fluctuations, leading to a distance independent stable orientation in the classical high temperature limit for all considered materials.


 Below, we consider the materials Gold (Au), Silicon (Si) and Polystyrene (PS). The medium outside the surface is either vacuum (V) or Bromobenzene (Br). The dielectric functions of these materials are provided in Appendix \ref{sec:param}.


\subsection{Stable orientations} 

\begin{figure}[h]
\includegraphics [width=0.95\columnwidth]{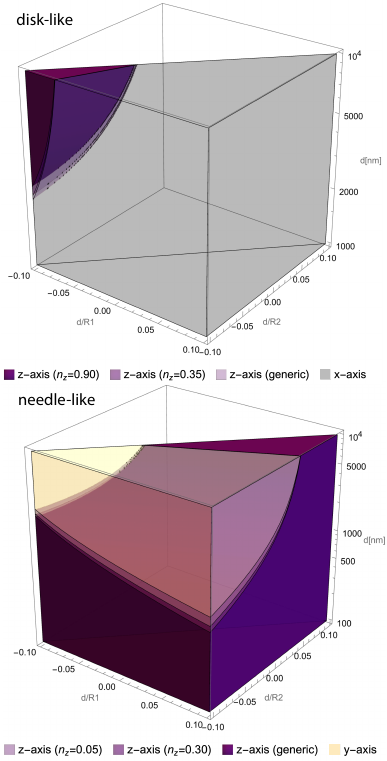} 
\caption{\label{fig:Au_stability} Stability diagram for Au surface in vacuum at $T=0$ K: For a flat surface ($d/R_1=d/R_2=0$) a disk-like (needle-like) particle positions itself with the symmetry axis parallel (perpendicular) to the surface at all distances $d$. For a curved surface, the stable direction can change with distance, depending on the radii of surface curvature.}
\end{figure}

\begin{figure}[t]
\includegraphics [width=0.95\columnwidth]{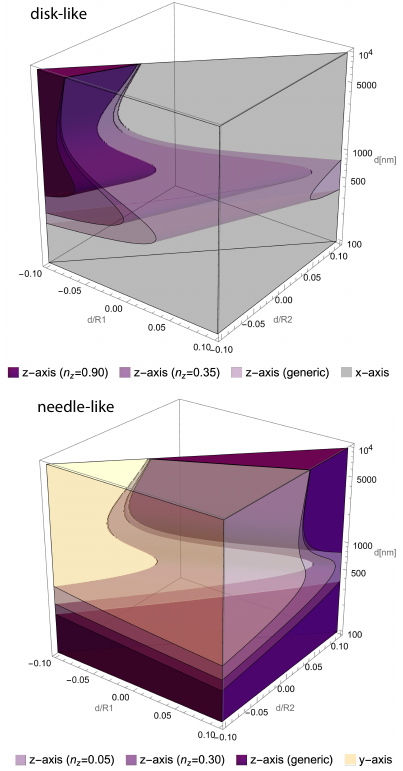} 
\caption{\label{fig:Au_in_Br_stability} Stability diagram for Au surface in Br at $T=0$ K: For a flat surface ($d/R_1=d/R_2=0$) 
the preferred orientation can change twice with distance, depending on the details of the particle's polarizability: A disk-like (needle-like) particle tends to position itself with the symmetry axis perpendicular (parallel) to the surface at intermediate distances $d$. For a curved surface, the one or both orientation switches can be eliminated, depending on curvature magnitude.}
\end{figure}

\begin{figure}[t]
\includegraphics [width=0.8\columnwidth]{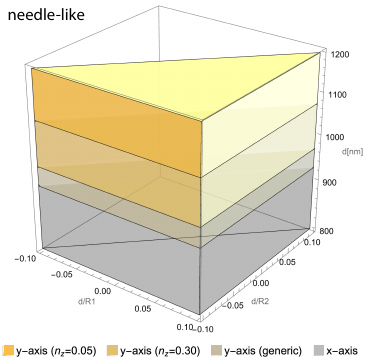} 
\caption{\label{fig:P_in_Br_stability} Stability diagram for polystyrene surface in Br at $T=0$ K: For both a flat surface ($d/R_1=d/R_2=0$) and a curved surface, a needle-like particle positions itself with the symmetry axis parallel to the surface but the preferred tangential direction depends on distance. A disk-like particle positions itself with the symmetry axis perpendicular to the surface at all distances.}
\end{figure}

Now we demonstrate how surface curvature affects stable particle orientations, using Eq.~(\ref{eq:U_orientation}). We consider two different types of particles: (1) a generic particle with a frequency independent polarization anisotropy $\sigma$, and (2) an ellipsoidal particle made of gold, silicon  or polystyrene, where the now frequency dependent polarization anisotropy is given by \cite{sihvola}
\begin{equation}
\label{eq:sigma_ellips}
\sigma(i\xi) = 
\frac{\epsilon_0 (\epsilon_1-\epsilon_0)^2 (1-3 \,n_z) V}{\left[\epsilon_0(1\!-\!n_z)+\epsilon_1 n_z \right] \left[\epsilon_0 (1\!+\!n_z)+\epsilon_1(1\!-\!n_z)\right]}\;,\end{equation}
where $\epsilon_1 \equiv \epsilon_1({\rm i}\,\xi)$, $\epsilon_0 \equiv \epsilon_0({\rm i}\,\xi)$, $V$ is the particle's volume and $n_z$ is the depolarization factor which is given by the eccentricity of the particle. For a disk-like particle $1/3<n_z<1$, and for a needle-like particle $0<n_z<1/3$.  
From Eq. (\ref{eq:sigma_ellips}), it is apparent that $\sigma<0$ for a disk-like particle, whereas $\sigma>0$ for a needle-like particle.
The surface is made of the same material as the particle. The surrounding medium is either vacuum or Bromobenzene. 
With $\sigma(i\xi)$ substituted in Eq.~(\ref{eq:U_orientation}) we determine the particle orientation with minimal $U_{o}(d,\theta,\phi)$. 
For $T=0$ K the stable axes as function of the surface-particle distance $d$ (between 100nm and 10$\mu$m) and the distance to curvature ratios are shown for Au in vacuum in Fig.~\ref{fig:Au_stability}, for Au in Br in Fig.~\ref{fig:Au_in_Br_stability} and for polystyrene in Br in Fig.~\ref{fig:P_in_Br_stability}. 
Here positive (negative) radii of curvature mean that the surface is curved towards (away from) the particle. Shown is only the region with $d/R_1 \ge d/R_2$ since the stability diagrams for $d/R_1 \le d/R_2$ is obtained by the replacement $z \to z$, $x \to y$, $y \to x$ for the stable axis.
For Au in vacuum the symmetry axis of a disk-like particle is oriented typically along the $x$-axis but there exists a small region where the particle orients along the $z$ axis above a curvature dependent distance $d$. A needle-like particle is oriented along the $z$-axis in proximity to the surface and aligned with the $y$-axis for larger distances. The switching distance depends (weakly) on the particle eccentricity via $n_z$. For Au in Br, the stability diagram is more complex. 
For a flat surface there can be two switches of stable orientation with distance for both disk- and needle-like particles. One or both of these switches can be eliminated due to surface curvature, depending on the relative magnitude of distance and radii of curvature. For polystyrene in vacuum a disk-like (needle-like) particle is {\it always} oriented along the $x$- or $y$-axis ($z$-axis), independent of distance and curvature. However, for polystyrene in Br, only a disk-like particle's orientation along the $z$-axis is independent of distance and curvature. A needle-like particle switches orientation at a curvature independent distance between $x$-axis close to the surface and $y$-axis at larger surface separation. The switching distance depends on the eccentricity of the particle. Finally, for Si  in either vacuum or Br, a needle-like (disk-lie particle) is {\it always} oriented along the $z$-axis ($x$- or $y$-axis), and hence no stability plot is shown for this case.

\begin{figure}[t]
\includegraphics [width=0.95\columnwidth]{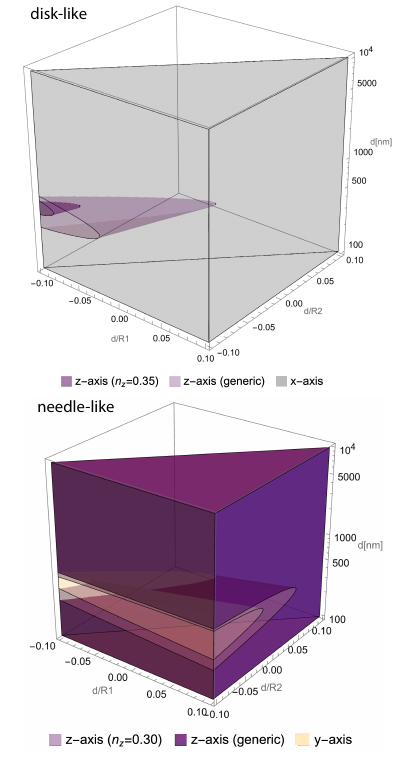} 
\caption{\label{fig:Au_in_Br_300_stability} Stability diagram for Au surface in Br at $T=300$ K: 
Compared to zero temperature, the generic stability regions ($x$-axis for a disk-like particle, $z$-axis for a needle-like particle) are enlarged by thermal fluctuations.
However, two changes of preferred orientation are still possible, depending on the curvature magnitude.}
\end{figure}

Next, we consider the effect of thermal fluctuations. It is instructive to consider first the 
high temperature limit corresponding to the $n=0$ term in Eq.~(\ref{derexpa}) only. The orientation dependent CP potential becomes in this limit
\begin{widetext}
\begin{equation}
\label{eq:CP_high_T}
U_{o} = -\frac{k_B T\sigma}{32\, \epsilon_0\, d^3} \,  \left[\left\{ \frac{\epsilon_1-\epsilon_0}{\epsilon_1+\epsilon_0} - \frac{(\epsilon_1-\epsilon_0)^2}{4(\epsilon_1+\epsilon_0)^2} \, \left(\frac{d}{R_1}+ \frac{d}{R_2}\right) \right\} \cos(2\theta) + \frac{(\epsilon_1-\epsilon_0)(3\epsilon_1+\epsilon_0)}{4 (\epsilon_1+\epsilon_0)^2} \, \left(\frac{d}{R_1}- \frac{d}{R_2}\right) \cos(2\phi) \sin^2 \theta \right] 
\end{equation}
\end{widetext}
where $\sigma$ and the permittivities are the static values here. Note that the potential is independent of the magnetic permittivities due to the decoupling of electric and magnetic modes in the static limit. From this result we find that for Au in vacuum or in Br, for Si in vacuum or in Br, and for polystyrene in vacuum a needle-like particle is always oriented along the $z$-axis and a disk-like particle along the $x$-axis ($y$-axis) for $1/R_1 > 1/R_2$ ($1/R_1 < 1/R_2$). However, for polystyrene in Br the stable orientation is the opposite: along the $z$-axis for a disk-like particle and along the $x$- or $y$-axis for a needle-like particle. This shows that thermal fluctuations remove the switches of stable orientation with distance. At room temperature ($T=300$K), the stable orientations are the same as in the high temperature limit with the exception of Au in Br. For this case the stability diagram is shown in Fig.~\ref{fig:Au_in_Br_300_stability}. Compared to the high temperature limit, regions with switched stable orientation proliferate at intermediate distances, and their size depends both on the magnitude of surface curvature and the eccentricity of the particle.

\section{Discussion} 
\label{sec:disc}
 
In this work, we have employed a MSE to derive curvature corrections to the CP interaction between small anisotropic particles and general magneto-dielectric surfaces. This approach represents a significant advancement over earlier derivative expansion methods, particularly in its ability to handle material effects and provide a direct curvature expansion without prior assumptions. We have derived an orientation-dependent CP potential, valid to linear order in $d/R_j$, where $d$ is the particle-surface distance and $R_j$ are the radii of curvature. 
Our results reveal complex and intriguing phenomena concerning the stable orientations of anisotropic nano-particles near curved surfaces. We show that the preferred particle orientation can depend on the distance $d$ from the surface, especially when the surrounding medium possesses specific optical properties. Crucially, we find that even small surface curvatures can induce or eliminate switches in stable particle orientation. These orientation switches are identified as a quantum effect, which diminishes with increasing thermal fluctuations, leading to a distance-independent stable orientation in the classical high-temperature limit. We illustrated these effects for various material combinations, including Gold, Silicon, and Polystyrene surfaces in vacuum or Bromobenzene. Our findings demonstrate that a strategic combination of material properties, surface curvature, and temperature can be effectively utilized to engineer and control the orientation of nano-particles in the vicinity of surfaces. This comprehensive theoretical framework provides valuable insights for the design and optimization of micro- and nano-mechanical devices leveraging these fundamental interactions.




\appendix

\section{Parameters of the dielectric functions}
\label{sec:param}


In our computations we used for the permittivities of the materials the  expressions 
\begin{eqnarray}
&&\!\!\!\!\!\!\!\!\!\!\!\!\epsilon_{\rm Au}({\rm i} \, \xi)= 1+\frac{\Omega_p^2}{\xi (\xi+\gamma)}+ \sum_j \frac{f_j}{\omega_j^2+g_j \xi+\xi^2}\;,\\
&&\!\!\!\!\!\!\!\!\!\!\!\!\epsilon_{\rm Si}({\rm i} \, \xi)= \epsilon_{\infty}^{(\rm Si)} + \frac{\epsilon^{(\rm Si)} _0- \epsilon^{(\rm Si)} _{\infty}}{ 1+ \xi^2/ \omega_{\rm UV}^2}\;,\\
&&\!\!\!\!\!\!\!\!\!\!\!\!\epsilon_{\rm polystyrene}({\rm i} \, \xi)=1+\sum_j \frac{f_j}{\omega_j^2+g_j \xi+\xi^2}\;,\\
&&\!\!\!\!\!\!\!\!\!\!\!\!\epsilon_{\rm bromobenzene}({\rm i} \, \xi)=1+ \frac{C_{\rm IR}}{1+\xi^2/\tilde{\omega}_{\rm IR}^2}+  \frac{C_{\rm UV}}{1+\xi^2/\tilde{\omega}_{\rm UV}^2}\;,
\end{eqnarray}
where $\Omega_p=9$ eV$/\hbar$,  $\gamma=0.035$ eV$/\hbar$ \cite{book2}, $\epsilon_{\infty}^{(\rm Si)} =1.035$, $\epsilon^{(\rm Si)} _0=11.87$, $ \omega_{\rm UV}= 4.34$ eV$/\hbar$, $C_{\rm IR}=2.967$, $C_{\rm UV}=1.335$, $\tilde{\omega}_{\rm IR}=0.360$ eV$/\hbar$,    $\tilde{\omega}_{\rm UV}=8.465$ eV$/\hbar$ \cite{bergstrom,hough},  and the oscillator parameters $\omega_j,f_j, g_j$ for Au and polystyrene are listed in Tables (\ref{tabAu}) and (\ref{tabpoly}) respectively.  

\begin{table}[h]
\begin{tabular}{ccc}
\hline
$\omega_j$ (eV$/\hbar$)   & $f_j$ (${\rm eV}^2$$/\hbar^2$)  & $g_j$ (eV$/\hbar$)\\
\hline
3.05    & 7.091     & 0.75\\
4.15    & 41.46     & 1.85 \\
5.4   & 2.7     & 1.0\\
8.5   & 154.7     & 7.0 \\
13.5    & 44.55     & 6.0 \\
21.5    & 309.6     & 9.0 \\
\hline
\end{tabular}
\caption{Oscillator parameteres for Au  \cite{decca2007} \label{tabAu}}
\end{table}

\begin{table}[h]
\begin{tabular}{ccc}
 \hline
 $\omega_j$ (eV$/\hbar$)  & $f_j$ (${\rm eV}^2$$/\hbar^2$)  &$g_j$ (eV$/\hbar$)\\
 \hline
6.35    & 14.6      & 0.65\\
14.0    & 96.9      & 5.0 \\
11.0    & 44.4            & 3.5\\
20.1    & 136.9     & 11.5 \\
\hline
\end{tabular}
\caption{Oscillator parameteres for polystyrene \cite{rev3} \label{tabpoly} }
\end{table}

\section{Free space Green tensors }
\label{sec:green}

The components of $6\times 6$ dimensional  free space Green tensor then are
\begin{equation}
\begin{aligned}
\mathbb{G}^{(EE)}_{\sigma,ij}({\bf r},{\bf r}')&=-\frac{1}{\kappa}\left(\frac{1}{\epsilon_\sigma} \frac{\partial^2}{\partial x_i \partial x'_j} + \mu_\sigma\, \kappa^2\, \delta_{ij} \right)\, g_\sigma( {\bf r}-{\bf r}')\;, \\
\mathbb{G}^{(HH)}_{\sigma,ij}({\bf r},{\bf r}' )&=-\frac{1}{\kappa}\left(\frac{1}{\mu_\sigma} \frac{\partial^2}{\partial x_i \partial x'_j} + \epsilon_\sigma\, \kappa^2\, \delta_{ij} \right)\,  g_\sigma( {\bf r}-{\bf r}')\;, \\
\mathbb{G}^{(HE)}_{\sigma,ij}({\bf r},{\bf r}')&=-\,\epsilon_{ijk} \frac{\partial}{\partial x_k}\,   g_\sigma( {\bf r}-{\bf r}') \;, \\
\mathbb{G}^{(EH)}_{\sigma,ij}({\bf r},{\bf r}')&= -\,\epsilon_{ijk} \frac{\partial}{\partial x'_k}\,  g_\sigma( {\bf r}-{\bf r}')\;,
\label{eq:App_G}
\end{aligned}
\end{equation}
where  $i,j\in \{x,y,z\}$ denote the spatial components, $\sigma$ labels the Green tensor in the surrounding space ($\sigma=0$) and the body ($\sigma=1$), $\epsilon_{ijk}$ is the Levi-Civita symbol, and the scalar Green function is
\begin{equation}
\label{eq:App_g}
g_\sigma( {\bf r}-{\bf r}')= \frac{e^{-\kappa \sqrt{\epsilon_\sigma \mu_\sigma} \vert {\bf r}-{\bf r}' \vert }}{4 \pi\,\vert {\bf r}-{\bf r}' \vert}\;.
\end{equation}

\end{document}